\documentclass[%
 reprint,
 amsmath,amssymb,
 aps,
]{revtex4-1}

\usepackage{graphicx}
\usepackage{dcolumn}
\usepackage{bm}

\usepackage{romannum}

\begin{document}


\title{Topological Landscape of Competing Charge Density Waves in 2H-NbSe\textsubscript{2}}

\author{Gyeongcheol Gye}
\affiliation{Center for Artificial Low Dimensional Electronic Systems, Institute for Basic Science (IBS), Pohang 37673, Republic of Korea}
\affiliation{Department of Physics, Pohang University of Science and Technology (POSTECH), Pohang 37673, Republic of Korea}
\author{Eunseok Oh}
\affiliation{Center for Artificial Low Dimensional Electronic Systems, Institute for Basic Science (IBS), Pohang 37673, Republic of Korea}
\affiliation{Department of Physics, Pohang University of Science and Technology (POSTECH), Pohang 37673, Republic of Korea}
\author{Han Woong Yeom}
\email{yeom@postech.ac.kr}
\affiliation{Center for Artificial Low Dimensional Electronic Systems, Institute for Basic Science (IBS), Pohang 37673, Republic of Korea}
\affiliation{Department of Physics, Pohang University of Science and Technology (POSTECH), Pohang 37673, Republic of Korea}

\date{\today}

\begin{abstract}
Despite decades of studies on charge density wave (CDW) of 2H-NbSe\textsubscript{2}, the origin of its incommensurate CDW ground state has not been understood. We discover that CDW of 2H-NbSe\textsubscript{2} is composed of two different, energetically competing, structures. The lateral heterostructures of two CDWs are entangled as topological excitations, which give rise to a CDW phase shift and the incommensuration without a conventional domain wall. A partially melt network of the topological excitations and their vertices explain an unusual landscape of domains. The unconventional topological role of competing phases disclosed here can be widely applied to various incommensuration or phase coexistence phenomena in materials. 


\end{abstract}
\maketitle

The competition of distinct periodicities or ordered phases are widely found in various of material systems \cite{feng2015,kundu2017,uehara1999,cheong2002}. 
Periodically ordered phases in spin \cite{geerken1982}, charge \cite{wilson1975,moncton1975,moncton1977}, and orbital degrees of freedom \cite{mori1998} have also interactions with underlying lattice periodicities, which result in commensuration-incommensuration phenomena and create domain-wall topological excitations \cite{bak1982,huang2017}. Notable examples are Frenkel-Kontorova solitons in one dimensional systems \cite{frenkel1938} and Moire structures in van der Waals layers \cite{alden2013,woods2014}. In most of the cases, competitions between ordered phases and those between them and underlying lattices have been separately discussed. However, here, we observe a unique case where those two phenomena, the competing phases and the incommensuration are topologically entangled to exhibit an unusual landscape of domains of coexisting phases. 

We pick up the charge density wave (CDW) in two dimensional (2D) materials, which is one of the most extensively investigated topics in condensed matter physics together with superconductivity \cite{gruner1988}. 
For most of 2D CDW phases \cite{withers1986}, ground states are commensurate with their atomic lattices, which would transit into incommensurate CDW phases with topological excitations, discommensurations or domain walls carrying a CDW phase shift \cite{mcmillan1976}, through thermal excitation \cite{fung1981}, doping \cite{yu2015, li2015} or pressure \cite{sipos2008, joe2014}.
Topological excitations are believed to have crucial roles not only in the superconductivity emerging with the commensurate CDW order suppressed \cite{sipos2008,yu2015,joe2014,li2015} but also in outstanding functionalities of CDW materials such as the ultrafast switching behavior \cite{stojchevska2014} and the memrister operation \cite{yoshida2015}. 
Nevertheless, atomic structures of domain walls are largely unknown \cite{cho2017} and thus naturally microscopic mechanisms for the emerging superconductivity and functionalities have been elusive.

In one of the most widely investigated model systems of 2D CDW, 2H-NbSe\textsubscript{2}, the intriguing nature of the incommensurate phase has become even a long term mystery.
2H-NbSe\textsubscript{2} undergoes a phase transition to the $(3 \times 3)$ CDW state at 33 K \cite{moncton1975,moncton1977}. 
While the microscopic origin of the transition is still under debates, the CDW ground state has been reported to be intriguingly incommensurate as confirmed by electron microscopy \cite{eaglesham1985}, nuclear magnetic resonance \cite{ghoshray2009}, and recent scanning tunneling microscopy (STM) \cite{Soumyanarayanan2013} experiments. However, none of these experiments identified the atomic structure of the discommensuration or the domain wall. 
Moreover, there is still ambiguity in the atomic structure of the commensurate CDW itself.  
In the early neutron scattering experiment, the CDW state was suggested to have a hexagonal symmetry \cite{moncton1977} but later the electron microscopy clarified the symmetry as an orthorhombic one \cite{eaglesham1985}. 
Three types of distinct CDW structures had been suggested in early theories \cite{walker1981, littlewood1982} and their atomic structures together with possible magnetic orders were discussed by a very recent density functional theory (DFT) calculation \cite{zheng2018}. 

\begin{figure*}
\includegraphics[scale=1]{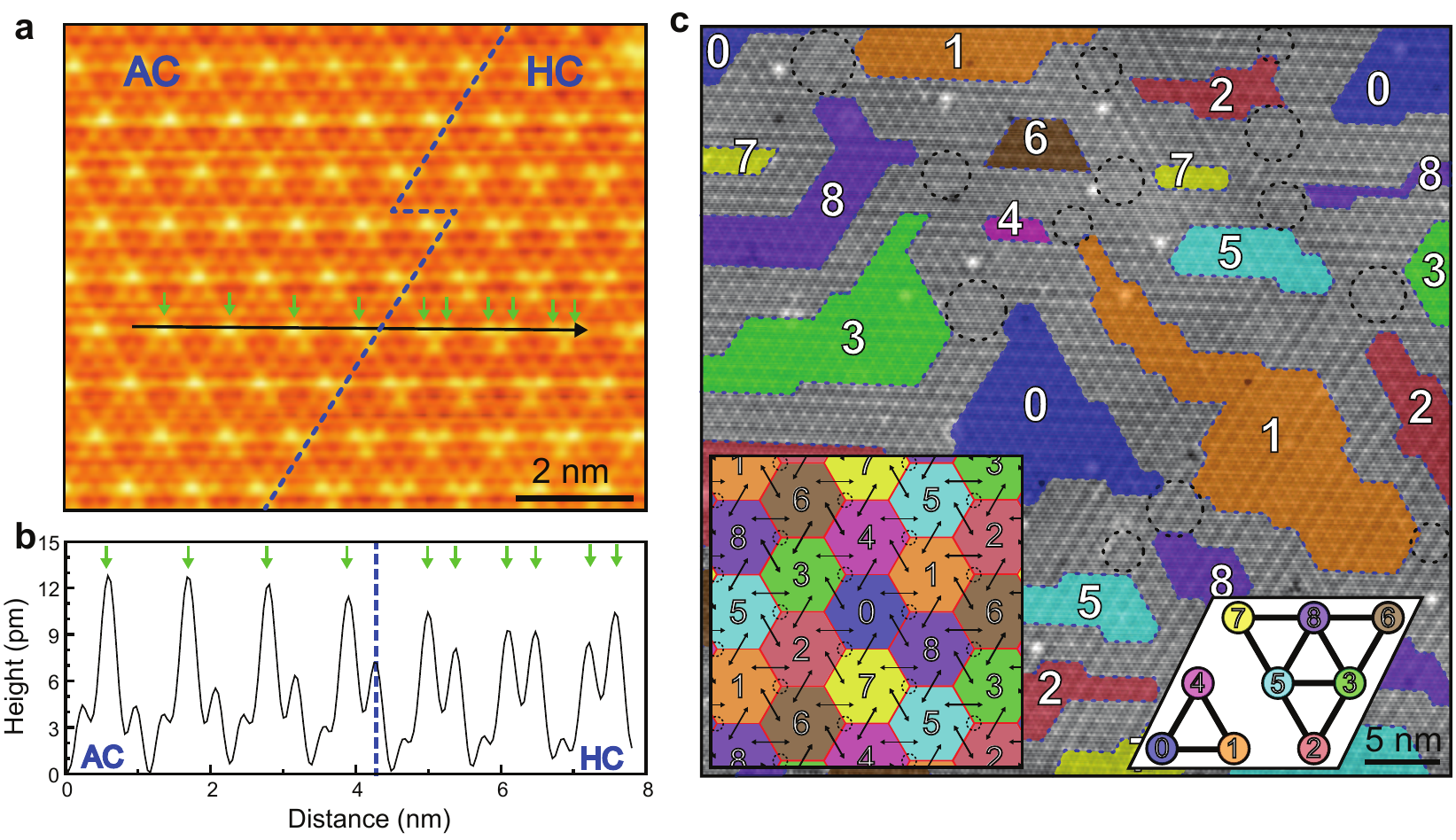}
\caption{\label{Fig1}  (a) STM topography of the coexisting domains in 2H-NbSe\textsubscript{2} at a sample bias of 100 mV. The blue dashed line separates two different domains. (b) Line profile along the black arrow in (a). Green arrows indicate the locations of bright topographic protrusions. (c) STM topography of HC CDW domains at 100 mV. Black dashed circles indicate the region of AC CDW dislocations as magnified in Fig. \ref{Fig4}(b). The colors and numbers indicates particular HC CDW domains among nine translationally degenerate domains. These domains have to be connected properly in the ideal domain network shown in the insets. Different domains are identified by their relative translation as defined in the right inset within one CDW unit cell. The black arrows represent the directions of CDW translations among HC CDW domains.}
\end{figure*}

In this work, we clarify the unconventional atomistic origin of the incommensurate CDW state in 2H-NbSe\textsubscript{2} using STM experiments and DFT calculations. 
The CDW state of 2H-NbSe\textsubscript{2} is divided by domains of two different CDW structures. 
Our DFT calculations and experimental results disclose atomic structures of the distinct CDW structures and their degenerate energies. 
The characteristic heterostructure of two CDW domains produces a phase shift for the discommensuration \cite{mcmillan1976}. Such a topological role of two competing phases explains the unusual domain distribution observed in the experiment. 
This reveals a distinct mechanism of the CDW discommensuration and a unique case of topologically entangled domain formation, which might be considered not only in CDW systems but also in a wide variety of materials with incommensuration phenomena and competing phases including Moire structures in van der Waals layers \cite{alden2013,woods2014}.

\begin{figure*}
\centering
\includegraphics[scale=1]{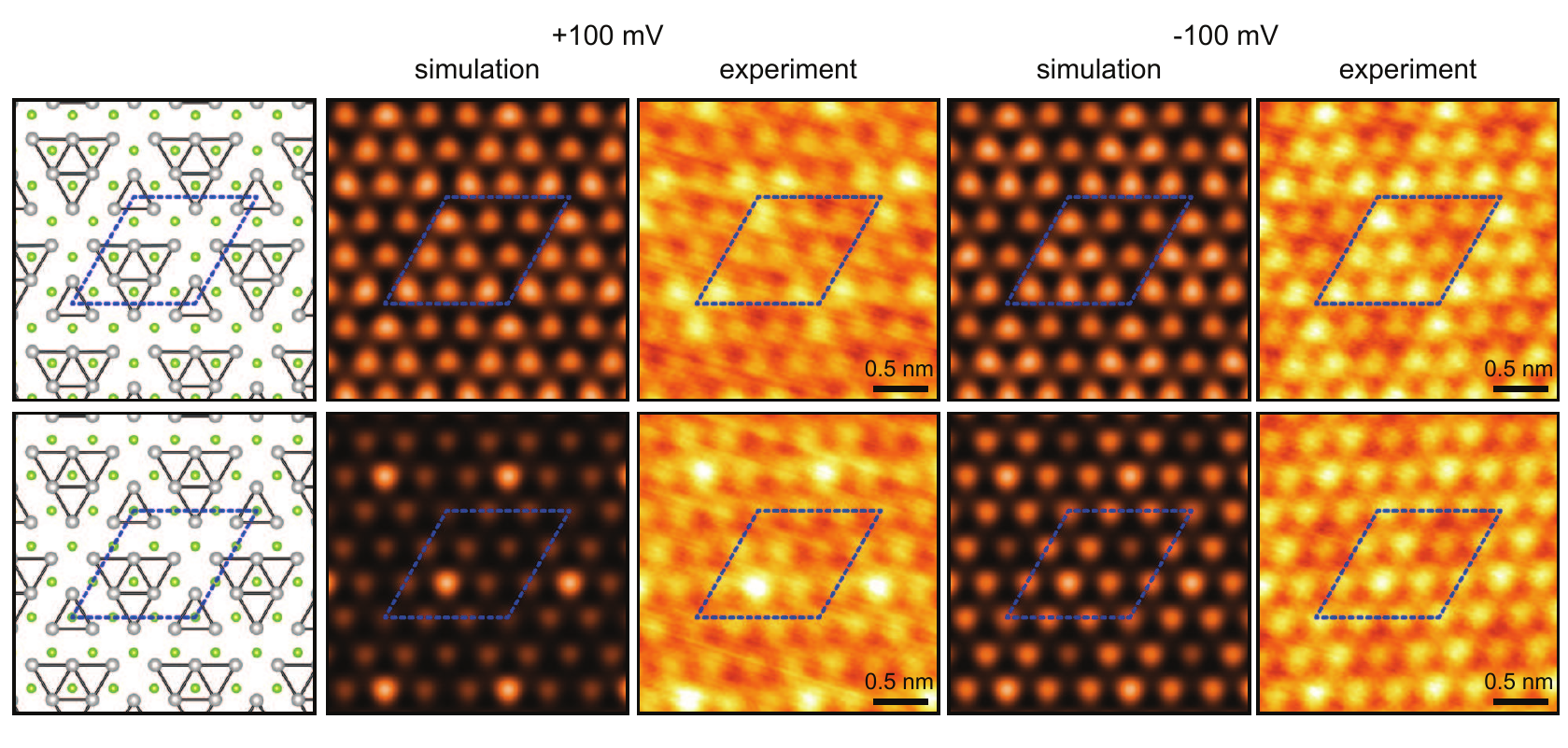}
\caption{\label{Fig2} Atomic structure models, simulated and experimental STM topographic images at two different biases of +100 and -100 mV of the HC (upper) and AC (lower) CCDW structures. The large gray (small green) balls represent Nb (Se) atoms in the structure model and the solid lines indicates short bonds for CDW distortions. The blue dashed lines indicate $(3 \times 3)$ CCDW unit cells.}
\end{figure*} 

Single crystal 2H-NbSe\textsubscript{2} is cleaved in high vacuum to obtain clean surfaces and transferred to the ultrahigh-vacuum chamber equipped with a commercial cryogenic STM operating at 4.3 K. 
STM images were obtained using the constant current mode. 
Our DFT calculations were performed using the Vienna ab initio simulation package \cite{kresse1996} with the projector augmented wave method \cite{blochl1994,kresse1999} and the Perdew-Burke-Ernzerhof exchange correlation functional \cite{perdew1996}. 
Single-layer structures of NbSe\textsubscript{2}, inserted between 10 $\AA$ vertical vacuum, were calculated with a $(6\times6\times1)$ k-point mesh and a plane wave cut-off energy of 366.5 eV. 
The residual force criterion to terminate atomic relaxation is 0.01 eV/$\AA$.

The STM topography of 2H-NbSe\textsubscript{2} shows clearly the coexistence of two different CDW structures [Fig. \ref{Fig1}(a)]. 
Hexagonally arrayed protrusions represent top Se atoms of NbSe\textsubscript{2} and the $(3 \times 3)$ CCDW order is clear with periodic enhancements of protrusions. 
Bright triangular protrusions or Se trimers are conspicuous in one of the CDW domains. The other domain exhibits a structurally distinct CDW state with a single bright Se protrusion per each $(3 \times 3)$ unit cell. 
Figure \ref{Fig1} (b) shows a detailed line profile across two CDW structures; the single big peak and two small ones for the CDW structure with monomer protrusions smoothly transform into two big peak structures which are parts of bright Se trimers. 
The coexisting CDW states is ubiquitous in our measurements over whole measured surfaces [Fig. \ref{Fig1}(c)], different cleavages, crystals and tunneling biases (see Fig. S1 in Supplemental Material \cite{Sup}). 
Moreover, the coexisting structures are noticeable in most of the STM images published previously although they have never been discussed explicitly \cite{Soumyanarayanan2013,dai2014}.
These CDW structures exhibit an intriguing domain structure as shown in Fig. \ref{Fig1}(c); the landscape of all-connected monomer and all-isolated trimer domains deviate certainly from a random heterogeneous domain mixture of competing phases. 
As revealed below, this unusual landscape has a topological origin.

We identify atomic structures of two CDW structures using STM images and DFT calculations [Fig. \ref{Fig2}]. 
As discussed in the previous literature, CDW maxima can locate on cation (Nb), anion (Se), or hollow sites to form different CDW structures \cite{walker1981,littlewood1982} (see Fig. S2 in Supplemental Material \cite{Sup}). 
The cation-centered (CC) structure has the hexagonal symmetry but the anion-centered (AC) and hollow-centered (HC) structures have the orthorhombic symmetry \cite{walker1981, littlewood1982}. The hexagonal symmetry is not consistent with the electron microscopy experiment \cite{eaglesham1985} and the CC structures are not favorable in our calculation and comparison to STM images \cite{Sup}. We thus just consider the latter two structures. 
In terms of Nb atom distortions relaxed in the corresponding DFT calculations, HC and AC CDW structures have similar units of a large triangle with six Nb atoms and a small triangle of a Nb trimer (Fig. \ref{Fig2}). 
However, the HC structure has Nb triangles centered on the hollow sites while the AC structure's triangles are centered on Se atoms. 

STM simulations of our $(3 \times 3)$  CDW structures reproduce well the local topographies. 
In the HC structure, the bright trimer protrusion of three Se atoms on the Nb trimer is well compared with the STM topography at 100 mV and the bright (dark) triangular protrusion of six (three) Se atoms is in good agreement with the experiment at -100 mV. 
In contrast, in the AC structure, the brightest protrusion made by the center Se atom of the Nb sextet agrees well with the STM topography at both biases.
The CDW ground state would be the HC structure which has a 13 (17) meV energy gain per $(3 \times 3)$ unit cell compared with the AC (CC) structure. 
A very recent DFT study, considering only HC and CC structures, also reported a consistent result \cite{zheng2018}. 

\begin{figure}
\includegraphics[scale=1]{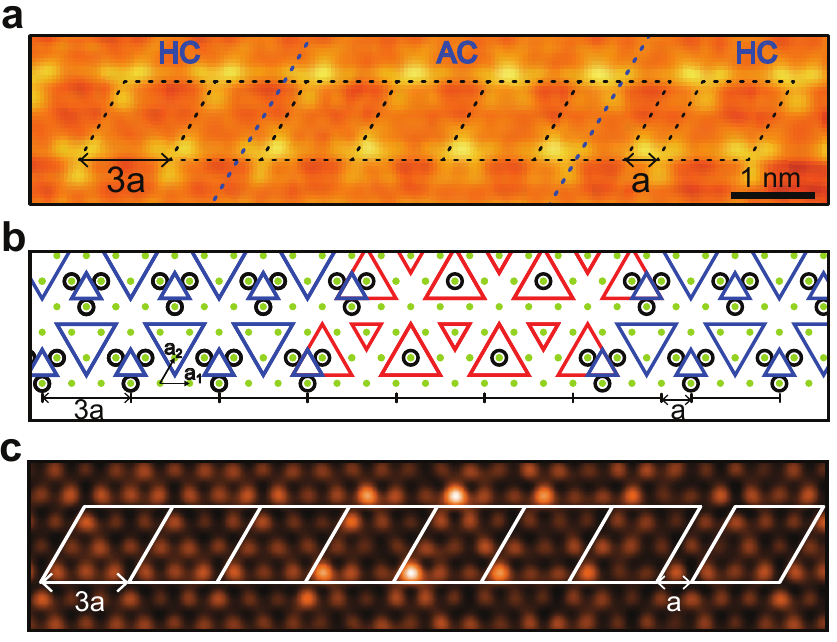}
\caption{\label{Fig3} (a) STM topography of a HC-AC-HC heterostructure in 2H-NbSe\textsubscript{2} at 100 mV. The blue dashed lines represent boundaries between HC and AC structures.  The black dashed lines indicate $(3 \times 3)$  CDW unit cells showing the a CDW translation between HC structures. (b) Schematic atomic and CDW configurations of the HC-AC-HC heterostructure. The blue and red lines indicate CDW distortions of Nb atoms in the HC and AC structures, respectively. The green dots represent Se atoms and black circles indicate the bright protrusions in STM. $\bold{a}_{1}$ and $\bold{a}_{2}$ are pristine lattice vectors to define the CDW translation, which is indicated for the HC structure at the bottom. (c) Simulated STM topography of the fully relaxed HC-AC-HC heterostructure at 100 mV. The white lines are drawn to indicate a CDW translation between HC structures.}
\end{figure}

With these CDW atomic structures, we can clarify the structural origin of the CDW discommensuration. As a matter of fact, a translation between HC and AC structures cannot be defined due to the absence of a common reference as shown in Figs. \ref{Fig1}(b) and \ref{Fig3} but one can define it between two HC domains sandwiching a stripe of an AC domain or vice versa.
Figure \ref{Fig3} shows such a structure observed by STM and its schematic atomic structure. 
This kind of structures is found all over the surface [see Fig. \ref{Fig1}(c)].
One of Se atoms nearby the Nb trimer (small blue triangle) in the HC structure and the Se atom at the center of a Nb sextet (large red triangle) in the AC structure are superimposed to connect the domains. 
This structure is fully relaxed in the DFT calculation and produces the STM image [Fig. \ref{Fig3}(c)], which 
is fully consistent with the observed one [Fig. \ref{Fig3}(a)].
The AC domain between two HC domains and two domain boundaries between the HC and AC structures together play the role of the discommensuration, which give rise to the translation of the HC CDW by a single unit cell of the pristine structure $\pm \bold{a_{1}}$. This corresponds to the phase shift of $(\mp \frac{2 \pi}{3} , 0, \pm \frac{2 \pi}{3})$ \cite{mcmillan1976,jacobs1982} in the HC CDW order parameter [Figs. \ref{Fig3}(a) and \ref{Fig3}(b)].

Among domain walls in CDW materials, those of 2H-NbSe\textsubscript{2} is unique. 
Most of CDW translations by discommensurations have been assumed between the same CDW structures with the translational degeneracy and is composed of well isolated broken CDW clusters. While there are very few examples of CDW domain walls with their structures determine, the recent case in 1T-TaS\textsubscript{2} shows such broken CDW clusters clearly \cite{cho2017}. 
In contrast to such an abrupt local change of structures, the present heterostructure for the CDW translation is a very smoothly connected structure \cite{cho2017}, making it hard to be identified. 
More contrastingly, while the domain walls of 1T-TaS\textsubscript{2} result in strongly localized in-gap states on the broken CDW clusters, the current HC-AC-HC heterostructure does not induce any strong electronic singularity (see Fig. S4 in Supplemental Material \cite{Sup}). 

In fact, the idea of degenerate CDW structures and their possible role as the discommensuration was introduced as early as 1982 for the case of 2H-TaSe\textsubscript{2} \cite{littlewood1982}. 
The main idea was that the two competing CDW structures in phase shifting heterostructures (HC-AC-HC in this work) can reduce the energy cost of conventional domain walls. 
While this idea didn't find clear experimental evidence in 2H-TaSe\textsubscript{2} yet but is fully consistent with what observed here. The only difference is that the previous theory considered only CC and AC domains, neglecting simply the HC structure  \cite{littlewood1982}. The free energy consideration of the previous work does not depend on any structural details of the phases involved \cite{littlewood1982} so that it can equally be applied to the present case of HC and AC competing structures \cite{Sup}.

\begin{figure}
\includegraphics[scale=1]{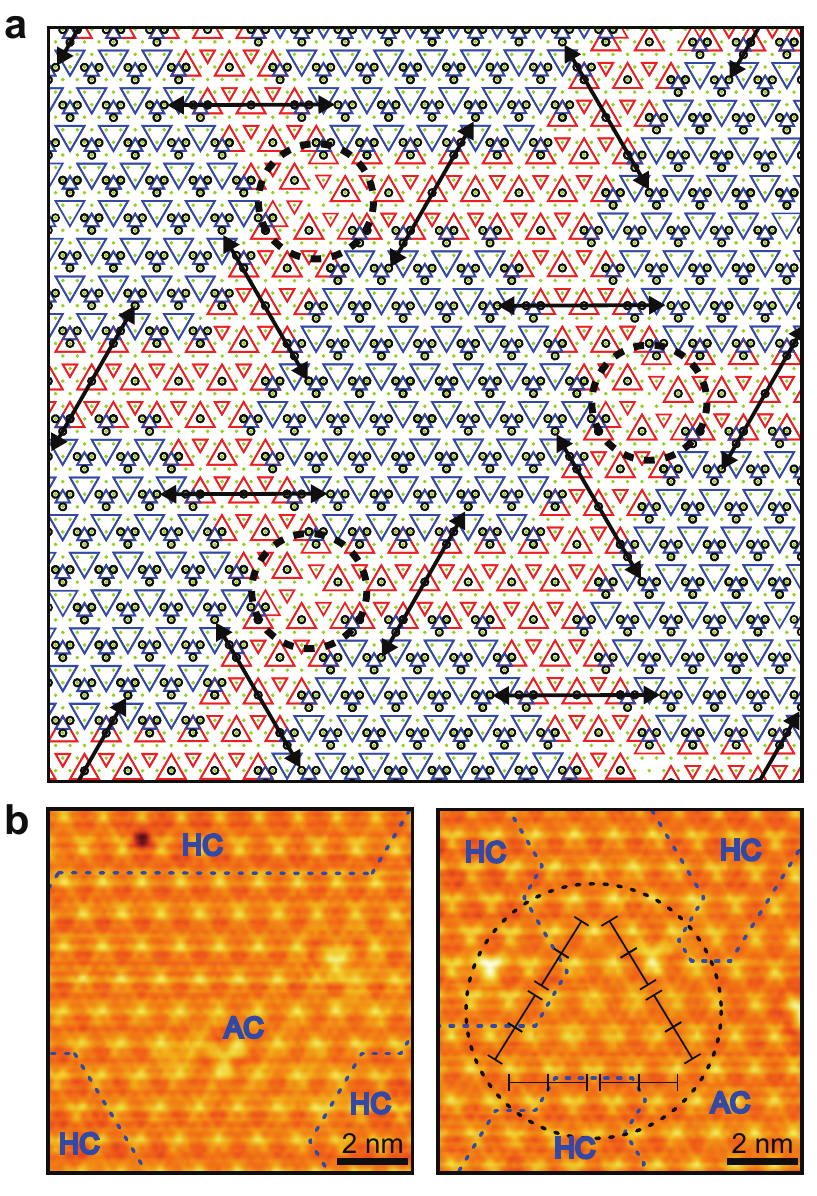}
\caption{\label{Fig4} (a) Atomistic schematics of an ideal honeycomb structure of the HC CDW domains and discommensurations  which are composed of type \Romannum{1} and \Romannum{2} vertices. Black dashed circle denotes the region of AC CDW dislocation and the black arrows the directions of CDW translations among the HC CDW domains. (b) STM topography of type \Romannum{1} (left) and \Romannum{2} (right) vertices where three AC CDW stripe domains merge. Black dashed circle indicates the region of AC CDW dislocations. Black solid lines represent the dislocations among the AC structures. 
}
\end{figure}

The discommensuration would ideally form a honeycomb lattice \cite{bak1982} as shown in the inset of Fig. \ref{Fig1}(c) (see the Supplemental Material for a detailed explanation \cite{Sup}). 
In the present system, this should correspond to hexagonal HC domains and the honeycomb network of the AC CDW stripes [Fig. \ref{Fig4}(a)]. 
 To form a honeycomb network, three AC CDW stripes should merge and the neighboring HC domains should have proper phase shifts [indicated by the domain numbering in Fig. \ref{Fig1}(c)]. 
 If the atomic structure of the HC-AC-HC discommensuration is put into this honeycomb network, the merging point of AC stripes is dictated to have two distinct structures; one with the dislocations (dashed circles) of AC CDW but the neighboring one without [Fig. \ref{Fig4}(a) and also see Fig. S5 Supplemental Material \cite{Sup}]. 
These correspond to a domain wall vortex and antivortex as dictated by the Z3 topology \cite{huang2017}. They are indeed observed in the STM topography; Fig. \ref{Fig4}(b) shows two representative merging points of AC stripes and only one of them shows the dislocations at its core (dashed circles). 
While the experimentally observed domain distribution in Fig. \ref{Fig1} is largely disordered, it can be closely mapped into the ideal honeycomb network \cite{Sup}. 
This network explains naturally why the HC and AC structures have distinct distributions of all isolated and all connected domains, respectively, and indicates the topological nature of the domain distribution as a partially melt network of topological excitations.

In this work, the long standing puzzle, the microscopic nature of the incommensurate CDW of 2H-NbSe\textsubscript{2} is thereby solved \cite{Sup}. 
The competition between HC and AC structures with a very small energy difference must be closely related to the quantum phase fluctuation observed very recently \cite{feng2015,kundu2017}. 
While the previous work suggested the 1Q phase as the source of the quantum fluctuation with the limited microscopic information, we very rarely observe the 1Q phase and its origin in the inhomogeneous strain \cite{Soumyanarayanan2013} is not easily compatible with the spontaneous quantum fluctuation.
The concept of topologically entangled competing phase and the discommensuration without a domain wall may find its application in various different materials systems with incommensuration phenomena.

The present work is supported by Institute for Basic Science (Grant No. IBS-R014-D1). The authors appreciate the discussion with C. Park.

\bibliographystyle{apsrev4-1}
\bibliography{ICDW}

\end{document}


\title{Supplemental Material for Topological Landscape of Competing Charge Density Waves in 2H-NbSe\textsubscript{2} }

\author{Gyeongcheol Gye}
\affiliation{Center for Artificial Low Dimensional Electronic Systems, Institute for Basic Science (IBS), Pohang 37673, Republic of Korea}
\affiliation{Department of Physics, Pohang University of Science and Technology (POSTECH), Pohang 37673, Republic of Korea}
\author{Eunseok Oh}
\affiliation{Center for Artificial Low Dimensional Electronic Systems, Institute for Basic Science (IBS), Pohang 37673, Republic of Korea}
\affiliation{Department of Physics, Pohang University of Science and Technology (POSTECH), Pohang 37673, Republic of Korea}
\author{Han Woong Yeom}
\affiliation{Center for Artificial Low Dimensional Electronic Systems, Institute for Basic Science (IBS), Pohang 37673, Republic of Korea}
\affiliation{Department of Physics, Pohang University of Science and Technology (POSTECH), Pohang 37673, Republic of Korea}
\date{\today}
\maketitle

\section{Various HC-AC-HC heterostructures and types of vertices}

The variation of the charge density, due to the CDW phase transition, is defined by $\delta \rho(\bold{r}) = Re[\sum_{i=1,2,3} e^{i \frac{1}{3} \bold{K}_{i} \cdot \bold{r} } \Psi_{i} (\bold{r})] $ with the three shortest reciprocal lattice vectors of the $(1 \times 1)$ unit cell $\bold{K}_{i}$, $i = 1, 2, 3$ as shown in Fig. S5(a). The component of the complex order parameter is written as $\Psi_{i} (\bold{r})=\Psi e^{i \theta_{i} (\bold{r})}$ \cite{mcmillan1975, jacobs1982}.
The phase shifts between order parameters of two HC structures in HC-AC-HC heterostructures are given by $(\mp \frac{2 \pi}{3} , 0, \pm \frac{2 \pi}{3})$ (left) $(\pm \frac{2 \pi}{3} , \mp \frac{2 \pi}{3}, 0)$ (mid) $(0 , \pm \frac{2 \pi}{3}, \mp \frac{2 \pi}{3})$ (right), respectively [Fig. S5(a)].

The AC CDW can propagate along the six directions of hexagonal lattice vectors. In the case of the HC-AC-HC heterostructure, the propagation of the AC structures in the direction of CDW translation vector between HC structures just broadens the AC CDW domains. 
Therefore, we only consider directions of the propagation of the AC CDW domain which are not parallel to HC CDW translation vector in the HC-AC-HC heterostructure and six types of the HC-AC-HC heterostructure are possible building blocks to construct two-dimensional domain structures, due to the three-fold symmetry [Fig. S5(a)]. 
The vertex, where two HC CDW domain edges with different CDW translation directions meet, is not allowed because there is a contradiction between the HC CDW translations [Fig. S5(c)]. 
Thus, two types of vertex which satisfy relations of CDW translations among three HC CDW domains are available [Figs. S5(d) and S5(e))]. 
The characteristic of the type \Romannum{1} vertex structure is triangle shaped HC CDW translation vectors (0-1-4 in Fig. 1(c) notation) around vertex while the type \Romannum{2} vertex structure has inverted triangle shaped HC CDW translation vectors (0-3-4 in Fig.  1(c) notation). In particular, because the stitching directions of HC-AC-HC structures and HC CDW translation directions in the type \Romannum{2} vertex are not compatible, the type \Romannum{2} vertex structure contains AC CDW translations in the core. Regardless of the direction of the AC CDW stripe, the direction of HC CDW translation vectors determine relative HC CDW domains and thus there are eight topologically equivalent vertex structures of type \Romannum{1} and type \Romannum{2}, respectively [Figs. S5(d) and S5(e)].
The closed HC CDW domain, constructed by type \Romannum{1} and type \Romannum{2} vertices, is unique and the idealized honeycomb structure of HC CDW domains composed of three each prototypical type \Romannum{1} and type \Romannum{2} vertices is represented in Fig. 4(a) of the main text. 

\begin{figure*}
\includegraphics[scale=1]{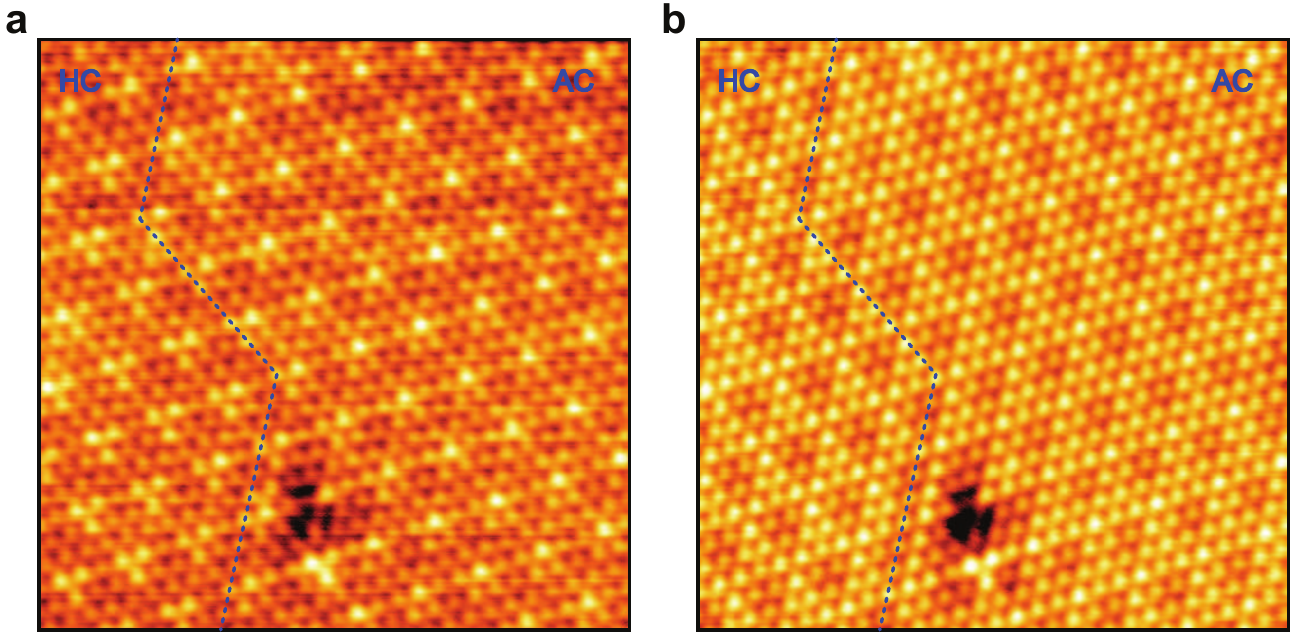}
\caption{\label{FigS1} 10 nm $\times$ 10 nm STM topography at 100 mV (a) and -100 mV (b). Blue dashed lines indicate the border of two domains. The two distinct domains are clearly observed regardless of bias.  }
\end{figure*}

\begin{figure*}
\includegraphics[scale=1]{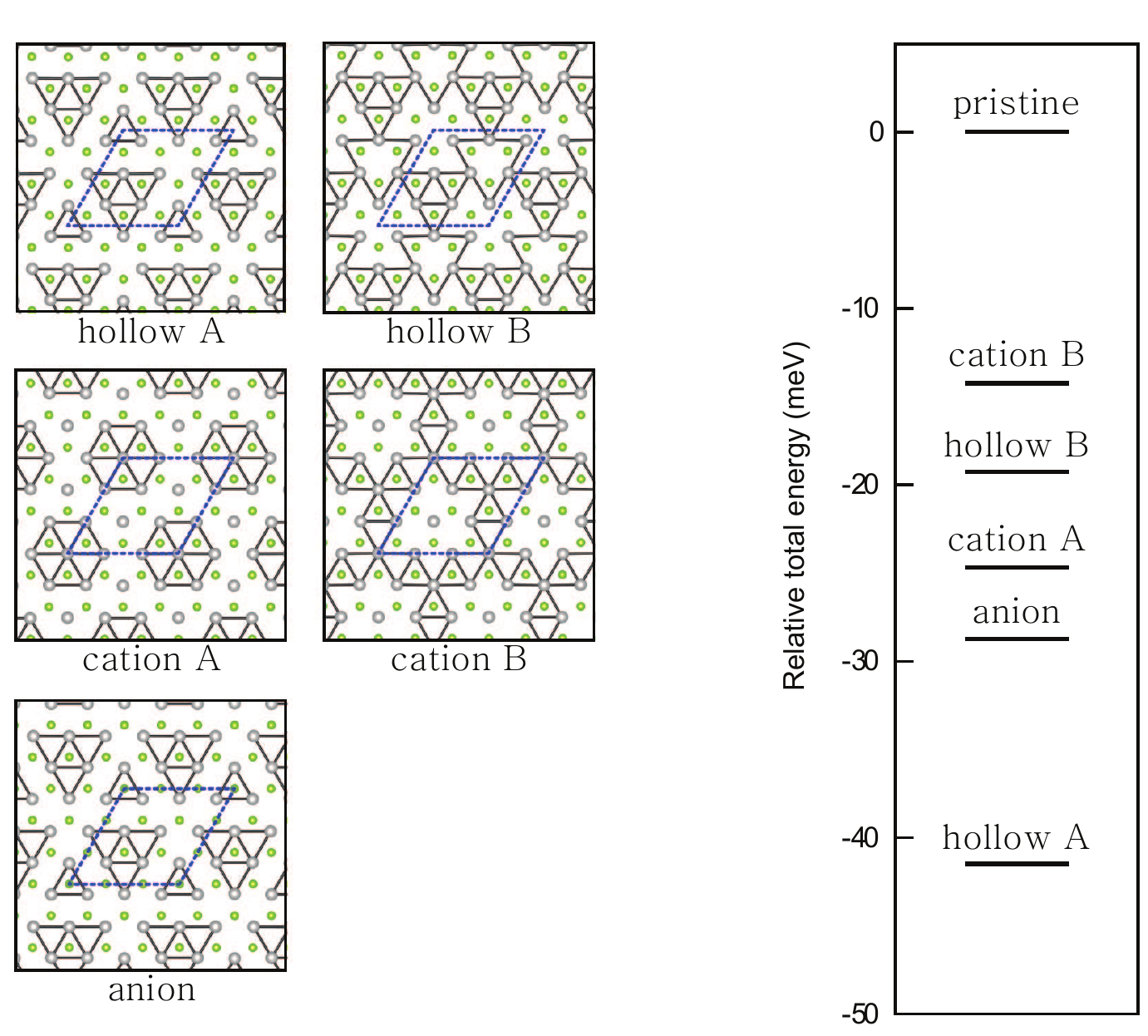}
\caption{\label{FigS3} Atomic structures and relative total energies of five CCDW structures of 2H-NbSe\textsubscript{2}. The CDW structures are categorized according to the center of the structure. Hollow A and anion centered structures correspond to HC and AC structures of the main text, respectively. The total energy of the pristine structure is set to zero. The energetics between HC and AC structure is not significantly altered even with the consideration of magnetic orders.}
\end{figure*}

\begin{figure*}
\includegraphics[scale=1]{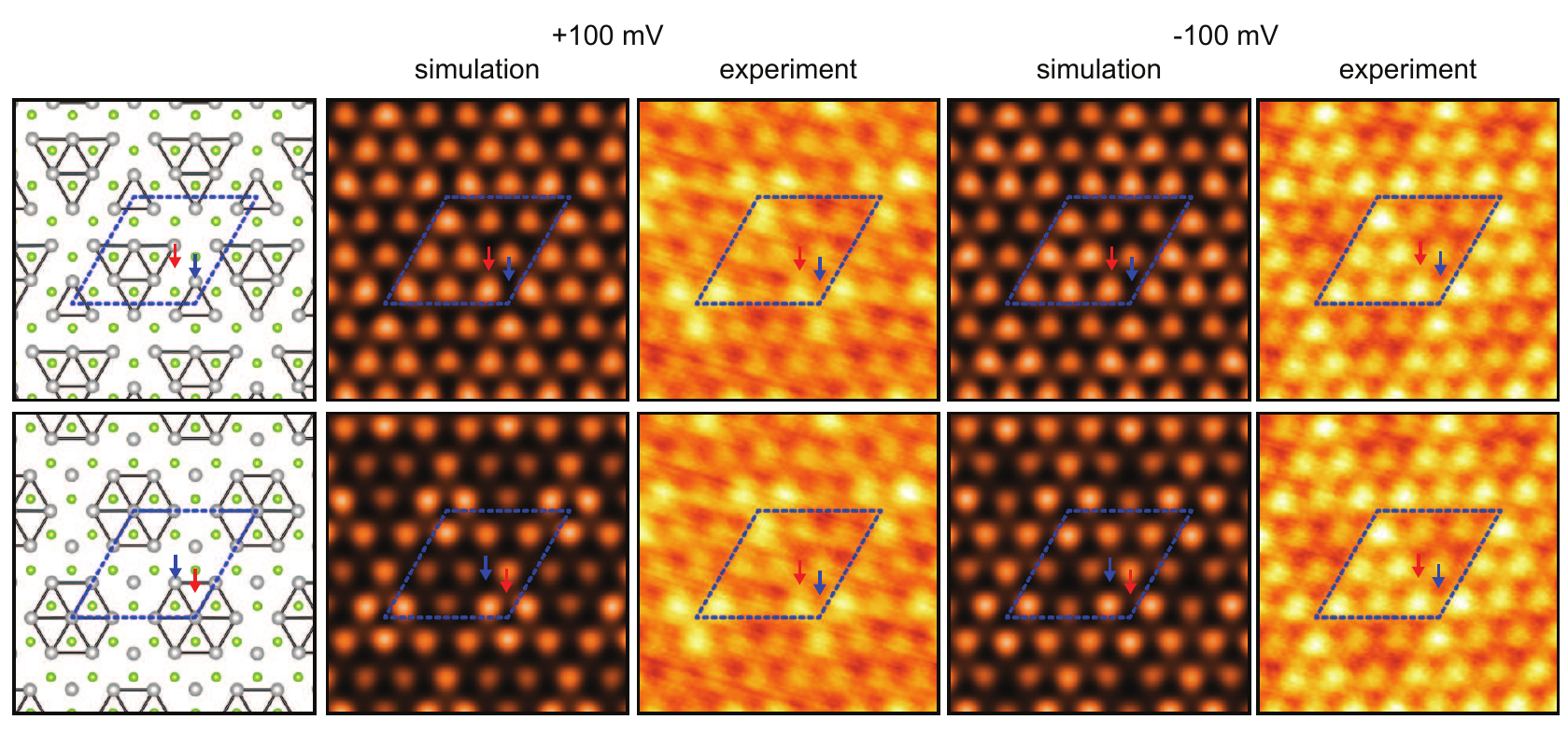}
\caption{\label{FigS2} Simulated STM topographic images of HC (upper) and cation A, corresponding to the CC structure, (lower) structures and experimental results at 100 mV and -100 mV. Red and blue arrows indicate hollow and Nb site, respectively. In both simulated images, hollow sites are brighter than Nb sites. Brightness of hollow and Nb sites in topography images of the HC structure reproduces well the experimental results but those of the CC structure are inconsistent with experimental ones.  }
\end{figure*}

\begin{figure*}
\includegraphics[scale=1]{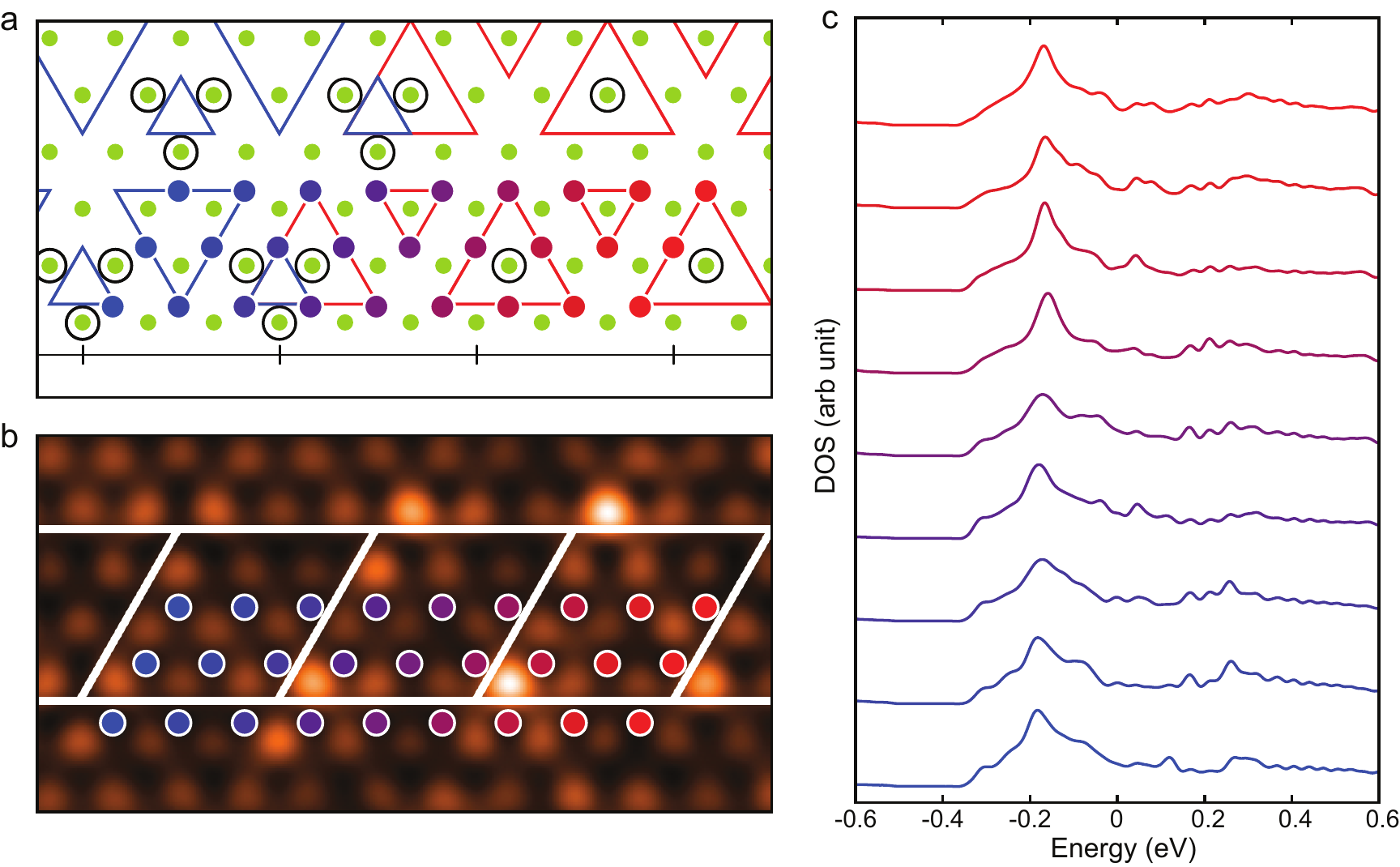}
\caption{\label{FigS4} (a) Schematics of the atomic structure near the HC-AC interface of the HC-AC-HC heterostructure as shown in Fig. 3(b) of the main text. Blue and red lines indicate CDW distortions of Nb atoms in the HC and AC structures, respectively. Colored (from blue to red) large dots represent Nb atoms near the interface of HC and AC structures.   Green dots represent Se atoms and black circles indicate bright Se atoms in the STM topography. (b) Simulated STM topography near the HC-AC interface as shown in Fig. 3(c) of the main text. (c) Averaged density of states of three Nb atoms by DFT calculations, corresponding to the same colors in (a) and (b). }
\end{figure*}

\begin{figure*}
\includegraphics[scale=1]{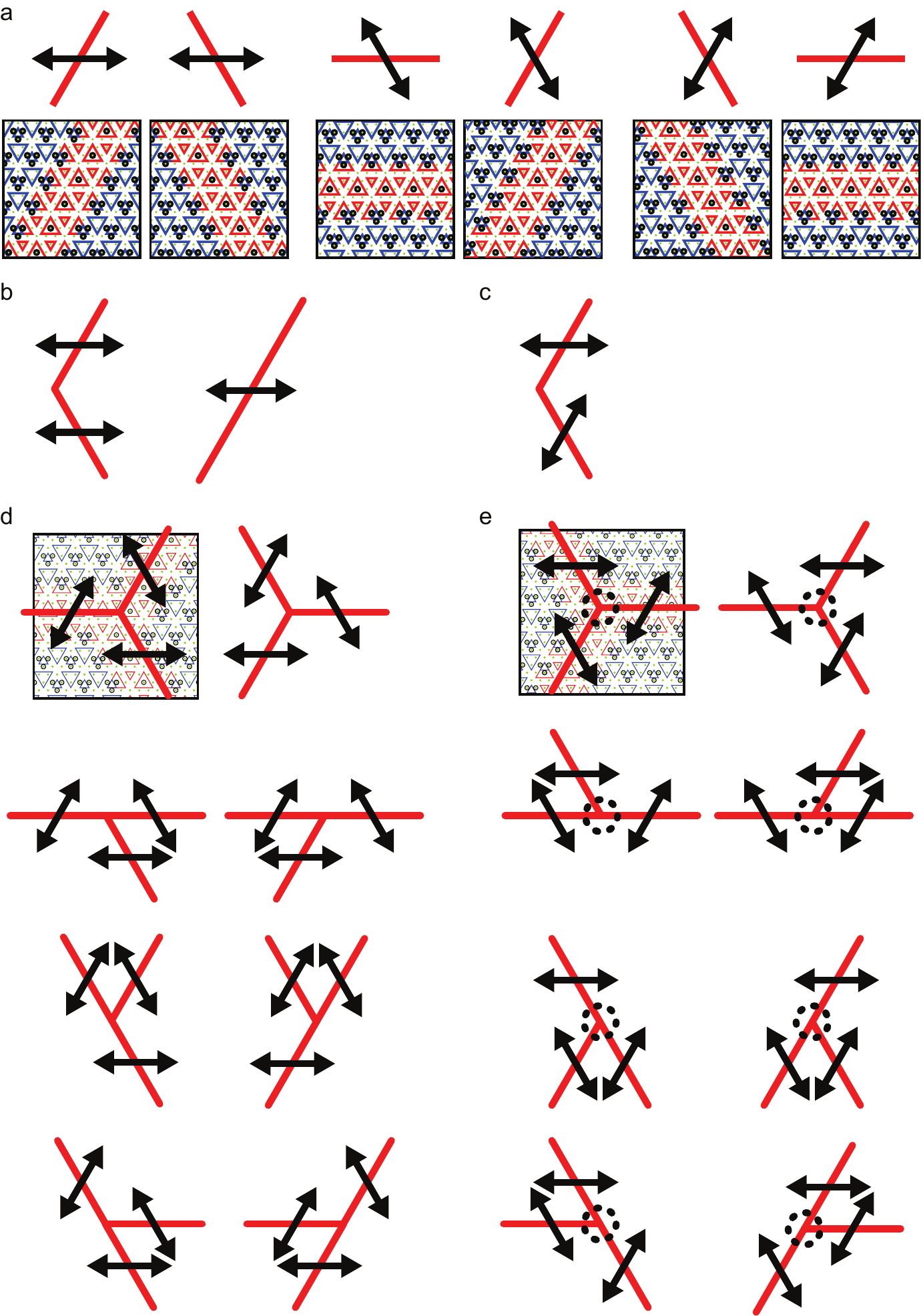}
\caption{\label{FigS5} (a) Line diagrams and schematic atomic structures of possible HC CDW translations due to the HC-AC-HC heterostructure. Black double head arrows indicate HC CDW translations and red lines represents the AC domains. (b) Line diagram for a vertex structure where edges of two HC domains meet (left). Line diagram for topologically equivalent structure of the vertex structure in the left (right). (c) An impossible vertex structure of two edges due to the contradiction of CDW translations between two HC domains. (d) Line diagrams for eight topologically equivalent structures of type \Romannum{1} vertex where three edges meet. (e) Line diagrams for eight topologically equivalent structures of type \Romannum{2} vertex.  }
\end{figure*}

\begin{figure*}
\includegraphics[scale=1]{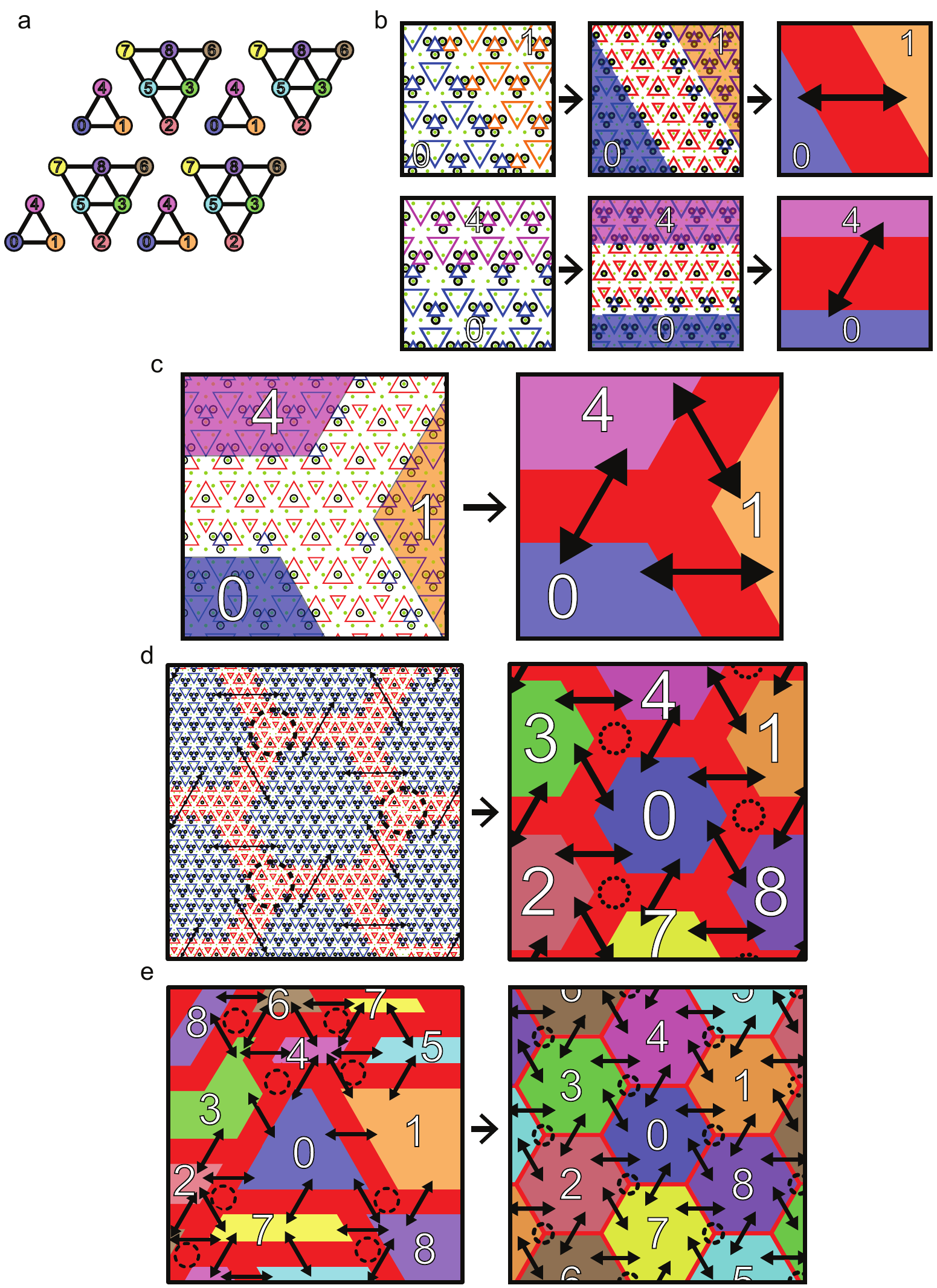}
\caption{\label{FigS6} (a) Numbers and colors indicate translational vectors relative to the 0 domains in the HC structure. (b) HC structures are colored to represent the relative translation between two adjacent domains with a conventional domain wall (left). HC-AC-HC heterostructures with two adjacent HC domains analogous to the left structures (mid). Schematic figures for the HC-AC-HC heterostructures (right). Red lines indicate AC domains. (c) A type I vertex with the colors and the numbers of three adjacent HC domains (left). Schematic figure for the type I vertex (right). (d) Honeycomb structure of HC-AC-HC heterostructures in Fig. 4(a) (left). Schematic figure for the honeycomb structure with the domain numbers and colors (right). (e) Schematic figure for disordered HC and AC domains (left). Schematic figure for the ideal honeycomb structure (right).  }
\end{figure*}

\begin{figure*}
\includegraphics[scale=1]{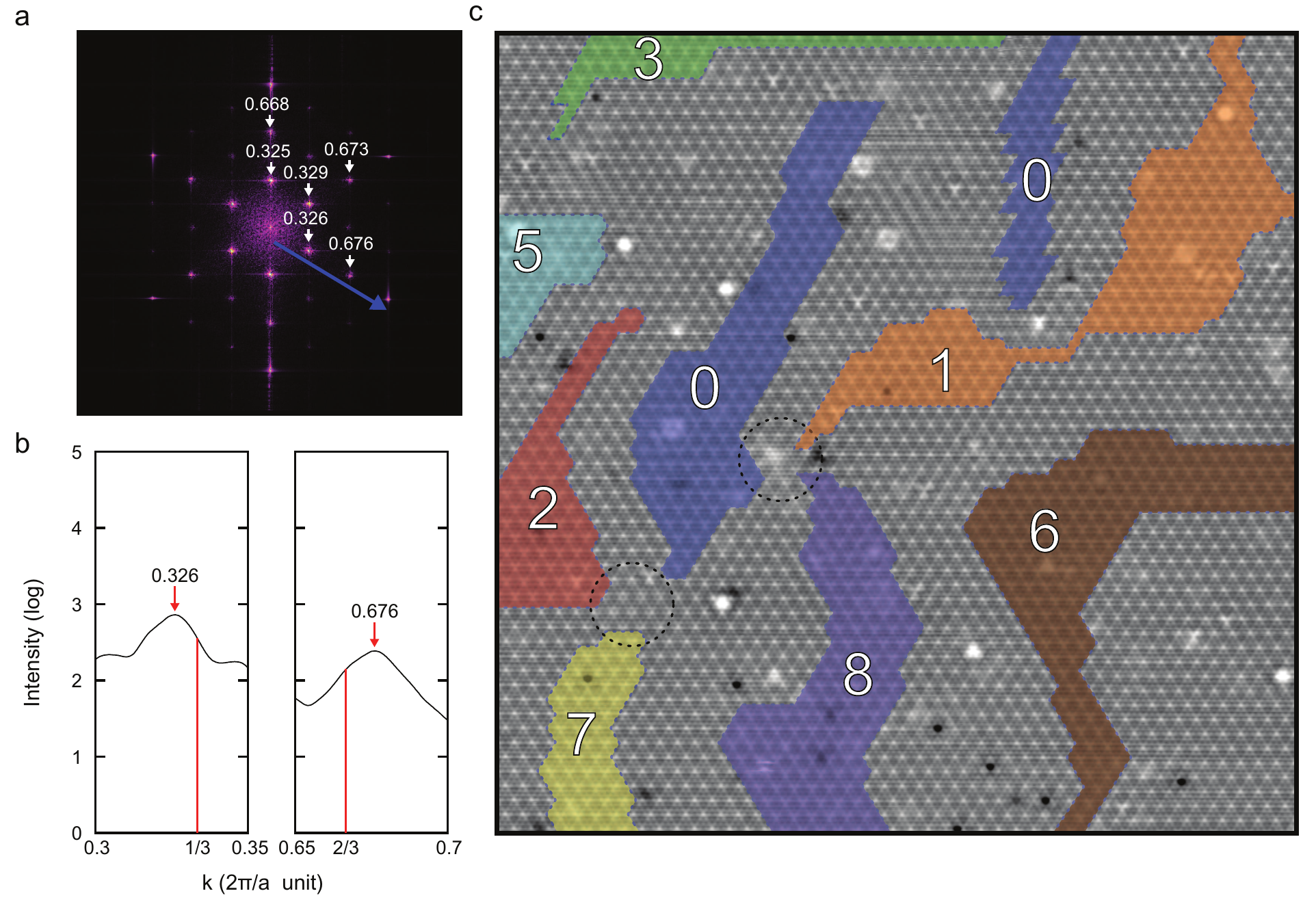}
\caption{\label{FigS7} (a) Fourier transform of the STM image in (c). (b) Peaks in the transformed image near 1/3 and 2/3 points parallel to the blue arrow in (a). $(1 \times 1)$ Bragg peak ($2 \pi /a$) is set to 1. (c) 50 nm $\times$ 50 nm STM topography at a sample bias of 100 mV.     }
\end{figure*}

\bibliographystyle{apsrev4-1}
\bibliography{SUP}